\documentclass[times,authoryear]{elsarticle}

\usepackage{jasr}
\usepackage{framed,multirow}

\usepackage{amssymb}
\usepackage{amsmath}
\usepackage{gensymb}
\usepackage{latexsym}
\usepackage{caption}
\usepackage{color,soul}
\usepackage{eso-pic, rotating, graphicx}

\usepackage[switch]{lineno}

\usepackage{url}
\usepackage{xcolor}
\definecolor{newcolor}{rgb}{.8,.349,.1}

\usepackage[citebordercolor=white]{hyperref}

\journal{Advances in Space Research}
\AddToShipoutPicture*{\put(575,700){\rotatebox{270}{\scalebox{1}{\copyright~2024. This manuscript version is made available under the CC-BY-NC-ND 4.0 license \url{https://creativecommons.org/licenses/by-nc-nd/4.0/}
}}}}

\begin{document}

\verso{Abdollah~Masoud~Darya~\textit{etal}}

\begin{frontmatter}

\title{Multi-instrument analysis of L-band amplitude scintillation observed over the Eastern Arabian Peninsula}%

\author[1]{Abdollah~Masoud~\snm{Darya}\corref{cor1}}
\cortext[cor1]{Corresponding author}
\ead{adarya@sharjah.ac.ae}
\author[2]{Muhammad~Mubasshir~\snm{Shaikh}}
\ead{mshaikh@sharjah.ac.ae}
\author[3]{Grzegorz~\snm{Nykiel}}
\ead{grzegorz.nykiel@pg.edu.pl}
\author[4]{Essam~\snm{Ghamry}}
\ead{essamgh@nriag.sci.eg}
\author[1]{Ilias~\snm{Fernini}}
\ead{ifernini@sharjah.ac.ae}

\affiliation[1]{organization={SAASST, University of Sharjah},
                city={Sharjah},
                country={United Arab Emirates}}

\affiliation[2]{organization={Department of Electrical Engineering, University of Sharjah},
                city={Sharjah},
                country={United Arab Emirates}}

\affiliation[3]{organization={Faculty of Civil and Environmental Engineering, Gdansk University of Technology},
                postcode={80-233 Gdansk},
                country={Poland}}

\affiliation[4]{organization={National Research Institute of Astronomy and Geophysics (NRIAG)},
                postcode={11421 Helwan},
                city={Cairo},
                country={Egypt}}


\begin{abstract}
The study of scintillation-causing ionospheric irregularities is important to mitigate their effects on satellite communications. It is also important due to the spatial and temporal variability of these irregularities, given that their characteristics differ from one region to another. This study investigates the spatial and temporal characteristics of L1 amplitude scintillation-causing ionospheric irregularities over the Eastern Arabian Peninsula during the ascending phase of solar cycle 25 (years 2020--2023). The temporal occurrences of weak and strong scintillation were separated by sunset, with weak scintillation observed predominantly pre-sunset during the winter solstice and strong scintillation observed mainly post-sunset during the autumnal equinox. Strong scintillation was much more pronounced in 2023 compared to the other three years, indicating a strong influence of solar activity. Spatially, weak-scintillation-causing irregularities exhibited a wide distribution in azimuth and elevation, while strong-scintillation-causing irregularities were concentrated southwards. The combined analysis of S4 and rate of total electron content index (ROTI) suggested that small-scale ionospheric irregularities were present in both pre- and post-sunset periods, while large-scale irregularities were only seen during the post-sunset period. Furthermore, the presence of southward traveling ionospheric disturbances (TIDs) during the 2023 autumnal equinox was confirmed with the total electron content anomaly ($\Delta\text{TEC}$), while the Ionospheric Bubble Index (IBI) provided by the Swarm mission was unable to confirm the presence of equatorial plasma bubbles during the same period. Observations from the FORMOSAT-7/COSMIC-2 mission indicated that strong-scintillation-causing irregularities were more prevalent under the F2-layer peak, while the weak-scintillation-causing irregularities were mostly observed at the E-layer, F2-layer, and above the F2-layer. This study aims to contribute insights into the behavior of scintillation-causing ionospheric irregularities in the region, with implications for future research during the peak of the 25th solar cycle.
\end{abstract}


\end{frontmatter}


\section{Introduction}
The scintillation of transionospheric radio signals greatly affects the accuracy of global navigation satellite systems (GNSS) and the quality-of-service of satellite communications, among other applications. Radio scintillation is caused by diffraction and scattering processes created by small-scale electron density perturbations or sharp electron density gradients \citep{tsagouri2023ionosphere}. It may also be caused by radio frequency interference \citep{10101815}. Studying the spatial and temporal distribution of scintillation-causing ionospheric irregularities is critical to further our understanding of this phenomenon.\par
Scintillation can be monitored through ground-based receivers with high or low sampling rates. The S4 index can be obtained from high sampling rate receivers, while the rate of total electron content index (ROTI) can be obtained from the numerous low sampling rate receivers. As the ionospheric irregularities vary in size, from centimeters to hundreds of kilometers \citep{ghobadi2020disentangling}, it is important to use different indices that can detect irregularities at different scale sizes. The S4 index, for instance, can detect irregularities in the order of a few hundred meters, whereas ROTI can detect irregularities of a few kilometers and above \citep{spogli2023stepping}. This provides an opportunity to study the occurrence of small-scale irregularities and compare it with the presence of large-scale irregularities. Recent studies conducted at low latitudes attempted to find the correlation between the S4 and ROTI indices. \citet{olwendo2018validation} found that ROTI values can be scaled to correlate well with low to moderate S4. Additionally, \citet{yang2016correlation} found that the correlation between the two values increases with solar activity.\par
While ground-based scintillation monitoring receivers can localize ionospheric irregularities with respect to elevation and azimuth or latitude and longitude, space-based radio occultation (RO) observations have the added advantage of locating irregularities height-wise, i.e., in terms of altitude. One recent mission providing RO S4 is the FORMOSAT-7/COSMIC-2 (F7/C2) mission. The F7/C2 mission follows the FORMOSAT-3/COSMIC mission, providing twice as many observations of the low-latitude region \citep{chen2021near}. It comprises six identical low Earth orbit (LEO) satellites operating at low inclination ($24\degree$). Recently, \citet{chen2021near} showed that the RO S4 provided by F7/C2 can detect equatorial plasma bubbles (EPBs). Furthermore, \citet{duann2023impact} used the F7/C2 S4 to study the ionospheric response to a geomagnetic storm.\par
The Swarm mission by the European Space Agency (ESA) detects EPBs through the Ionospheric Bubble Index (IBI) by correlating disturbances in the magnetic field and electron density \citep{park2013ionospheric}. The Swarm constellation consists of three satellites: Alpha, Bravo, and Charlie, launched on 22 November 2013. All three have a circular near-polar orbit ($87.5\degree$ inclination). Alpha and Charlie are flying side-by-side, separated by $1.4\degree$ in longitude with an altitude of about 470 km, whereas Bravo orbits at roughly 520 km \citep{friis2008Swarm}. Several studies have used the IBI onboard Swarm spacecraft. \citet{wan2018climatology} presented a statistical study of equatorial plasma depletions during pre- and post-midnight hours. \citet{hussien2021statistical} investigated the characteristics of plasma bubbles during different types of geomagnetic storms. More recently, \citet{reddy2023predicting} used Machine Learning to predict the IBI onboard Swarm.\par
Studying ionosphere irregularities can either be done using ground-based \citep{olwendo2021monthly} or space-based instruments \citep{chang2023variability}, or both \citep{miller2008multi,tilahun2020evening}. One of the first papers to propose using multi-instrument analysis for the localization of scintillation-causing ionosphere irregularities was \citet{miller2008multi}. They used field-aligned airglow images in addition to RO observations from the FORMOSAT-3/COSMIC mission, and ground-based scintillation monitors for the VHF and L-bands. Their study was focused on the equatorial American sector for the period of 7 September–-27 October 2006, corresponding to the autumnal equinox. They detected irregularities during pre-midnight and post-midnight ranging between 175--375 km in altitude. They also noted that higher altitude irregularities are uncommon during quiet geomagnetic conditions.\par
More recently, \citet{tilahun2020evening} used a combination of ground-based L-band and VHF scintillation monitoring receivers, in addition to true height of the peak electron density (hmF2) and the critical frequency of the F2 layer (FoF2) observations from the FORMOSAT-3/COSMIC mission and ground-based ionosonde to study ionospheric irregularities in the east Africa region during the summer solstice and autumnal equinox of 2014. He pointed out that strong scintillation was observed from 19:00 LT to midnight.\par
This work uses ground and space-based instruments to study the spatial and temporal distribution of scintillation-causing ionospheric irregularities over the Eastern Arabian Peninsula region for the ascending stage of solar cycle 25. This study is important for three main reasons. First, this work aims to fill the research gap in ionospheric studies over the Arabian Peninsula region. Second, ionospheric scintillation is the largest source of GNSS positioning errors in low latitude regions \citep{dubey2006ionospheric}. It also negatively impacts satellite communications. Mitigating these factors is important for the socioeconomic development of the countries in the region \citep{alameri2023Leveraging}. Third, understanding the behavior of scintillation-causing irregularities over the Arabian Peninsula, which is located at the northern crest of the equatorial ionization anomaly (EIA), may also aid in understanding the behavior of irregularities over regions at similar geomagnetic latitudes \citep{darya2019longitudinal}.\par
The paper is divided into the following sections: the methodology followed in this paper is described in Section \ref{Meth}. Next, analysis results are discussed in Section \ref{RnD}, and the conclusion is presented in Section \ref{Conc}.\par

\section{Methodology}\label{Meth}
This work utilizes observations from a multi-constellation GNSS PolaRx5S receiver (SHJ1) located at Sharjah, United Arab Emirates (Geog.: $25.28\degree$N, $55.46\degree$E; Geom.: $19.12\degree$N, $131.09\degree$E). This region is located at the northern crest of the EIA. The main goal of this study is to analyze the temporal (diurnal and seasonal) and spatial distribution of ionospheric irregularities that cause weak and strong L1 amplitude scintillation over the Eastern Arabian Peninsula, during the ascending phase of solar cycle 25. Hence, the selected study period spanned the years 2020--2023. Each year was divided into four seasons, each covering three months. The vernal equinox (VE) covers the months of February--April, the summer solstice (SS) covers the months of May--July, the autumnal equinox (AE) covers the months of August--October, and the winter solstice (WS) covers the months of November, December, and January.\par
Different ground and space-based instruments were used in this work, including:
\begin{itemize}
    \item Ground-based multi-constellation GNSS receiver.
    \item GNSS-RO measurements from F7/C2.
    \item In situ observations of equatorial plasma bubbles from Swarm Alpha and Bravo.
    \item Ground-based ionosonde measurements.
\end{itemize}
The following subsections will describe the data products used from these different instruments.\par

\subsection{S4}
Amplitude scintillation is observed through the S4 index. The total S4 index ($\text{S4}_\text{total}$) is defined as the standard deviation of the 50 Hz signal intensity (\textit{SI}) normalized to the average signal intensity over 60 seconds. The corrected S4 index eliminates ambient noise ($\text{S4}_\text{noise}$) from the total S4 index and is represented as \citep{van1993ionospheric}
\begin{equation}
\begin{aligned}
\label{eq1}
\text{S4}&=\sqrt{\text{S4}_\text{total}-\text{S4}_\text{noise}}\\
\text{S4}&=\sqrt{\frac{\langle\textit{SI}^2\rangle-\langle\textit{SI}\rangle^2}{\langle\textit{SI}\rangle^2}-\frac{100}{\textit{SNR}}\left[1+\frac{500}{19~\textit{SNR}}\right]},
\end{aligned}
\end{equation}
where \textit{SNR} is the signal-to-noise ratio of the signal. Note that for some instances, particularly for low total S4 values, the corrected S4 value was negative. These negative values were set to zero.\par
The 5-minute averaged L1 S4 observations from the GPS, Galileo, and BeiDou constellations were considered in this work. For the BeiDou constellation, only the MEO satellites were considered, since the low elevation of the GEO and IGSO BeiDou satellites, as observed from SHJ1, cause them to show elevated levels of S4 \citep{darya2022mapping}.
The following pre-processing steps were taken:
\begin{itemize}
    \item All observations with S4 $<0.1$ were considered as noise and excluded from this work since this work aims to understand the temporal and spatial distribution of amplitude scintillation-causing ionospheric irregularities.
    \item An elevation threshold of $30\degree$ was set to avoid multipath effects.
    \item All observations from geomagnetically active periods, where DST $<-50$ nT, were discarded.%
\end{itemize}
Two levels of S4 are considered in this work: $0.1<$ S4 $\leq 0.3$ represents weak scintillation, while S4 $>0.3$ represents strong scintillation \citep{spogli2013assessing}.\par
The S4 index was used in this work to detect ionospheric irregularities with scale sizes below the Fresnel's scale for L-band signals, in the order of a few hundred meters.

\subsection{ROTI}
ROTI is a commonly used GNSS-based index for detecting and measuring ionospheric irregularities \citep{pi1997monitoring}. It is defined as the standard deviation of the rate of total electron content (TEC) and can be used to study ionospheric irregularities \citep{LISSA2023}. ROTI, for the epoch $t$, can be calculated based on the Rate of TEC (ROT) values and selected time window $N$
\begin{equation}
\label{eq2}
\text{ROTI}=\sqrt{\frac{1}{N} \sum_{i=t-N}^{t} \left(\text{ROT}(i)-\overline{\text{ROT}}\right)^2},
\end{equation}
where ROT is the TEC differences between two epochs divided by the sampling interval ($\Delta t$)
\begin{equation}
\label{eq3}
\text{ROT}(t)=\frac{\text{TEC}(t)-\text{TEC}(t-1)}{\Delta t}.
\end{equation}
Both ROTI and ROT are expressed in TECU/min, where 1 TECU $= 10^{16}$ electrons$/m^2$. The ROTI in this work is based on 30-second ($\Delta t$) GPS observations and a time window ($N$) equal to 5 minutes.\par
The ROTI index was used in this work to detect ionospheric irregularities with scale sizes in the order of a few kilometers or more.\par

\subsection{$\Delta\text{TEC}$}
This work also analyzed the TEC variability, which can be used to detect traveling ionospheric disturbances (TID) \citep{olwendo2023simultaneous}. For that purpose, band-pass filtering was used to obtain the TEC anomaly in TECU ($\Delta\text{TEC}$). $\Delta\text{TEC}$ is calculated as a difference between a moving average with two different window sizes
\begin{equation}
\label{eq4}
\Delta\text{TEC}(t)=\frac{1}{T_1}\sum_{n=t-0.5\cdot T_1}^{t+0.5\cdot T_1} \text{TEC}(n) - \frac{1}{T_2}\sum_{n=t-0.5\cdot T_2}^{t+0.5\cdot T_2} \text{TEC}(n),
\end{equation}
where $T_1$ and $T_2$ are the sizes of time windows. This work used 15-minute/$\Delta t$ and 60-minute/$\Delta t$ windows, respectively \citep{nykiel2024large}.\par
The $\Delta\text{TEC}$ index was utilized in this work to investigate the presence of traveling ionospheric disturbances (TIDs) \citep{olwendo2023simultaneous} in the local ionosphere. TIDs are electron density fluctuations that propagate as waves in the ionosphere and can induce scintillations in trans-ionospheric radio wave signals \citep{priyadarshi2018polar}.

\subsection{F7/C2}
The Tri-GNSS Radio-occultation System (TGRS) on board the F7/C2 satellites measures the S4 index for all GPS and GLONASS satellites it tracks using a sliding window of 10 seconds of data, and the output is provided at 1-second intervals. The level 2 scintillation product (\emph{scnLv2}) was considered in this work to localize the ionospheric irregularities with respect to altitude.\par
The following parameters were extracted from the publicly available F7/C2 data:
\begin{itemize}
    \item From \emph{Global Data}, the start time of the data arc in GPS seconds (\emph{startTime}).
    \item From \emph{Profile Data}, time since start of scintillation event (\emph{time}), amplitude scintillation on L1 (\emph{S4\_L1}, and ECI coordinates of the F7/C2 (\emph{x\_LEO, y\_LEO, z\_LEO}) and GNSS satellites (\emph{x\_GPS, y\_GPS, z\_GPS}).
\end{itemize}
It is noted that the time for each observation is simply the sum of \emph{startTime} and \emph{time}.\par
By assuming, for each observation, a straight link (line one) between the F7/C2 and GNSS satellites (with known coordinates), the tangent point (unknown) is the point along the propagation link that is at the shortest distance from the center of the Earth (line two) \citep{hsu2018assessment}. The tangent point can then be found by exploiting the orthogonal nature of the two lines.\par
Since this work only considers S4 $>0.1$, all observations with S4 $<0.1$ were discarded. Furthermore, several thresholds were put in place to ensure that the analyzed data corresponds to the region and time of interest, namely: geographical latitude $\in[20\degree,28\degree]$, geographical longitude $\in[47\degree,63\degree]$, AE post-sunset $\in[19,23]$ local time (LT) hours, and WS pre-sunset $\in[14,18]$ LT hours. The choice of parameters will become apparent in Section \ref{RnD}. Similar to the ground-based S4 observation, the WS is considered to be the months of November, December, and January. However, due to the unavailability of \emph{scnLv2} observations for January 2022, January 2023 observations were considered instead. For the AE, observations from August--October were considered.\par

\subsection{Ionosonde}
Scaled observations were taken from a nearby Lowell Digisonde (DH224) located at Al-Dhafra Air Base (Geog.: $24.24\degree$N, $54.58\degree$E; Geom.: $18.12\degree$N, $130.11\degree$E), Abu Dhabi, United Arab Emirates. The auto-scaled observations of the hmF2 are provided at 7.5-minute intervals \citep{reinisch2011global}. The Automatic Real-Time Ionogram Scaler With True Height (ARTIST) program \citep{reinisch1983automatic} performed the automatic scaling. A confidence threshold of $\geq 95\%$ was utilized in this work, i.e., observations with confidence under $95\%$ were discarded to ensure the reliability of the auto-scaled observations \citep{themens2022artist}.\par
The hmF2 observations were used in this work to relate the \emph{scnLv2} observations to the peak density height.\par

\subsection{Swarm}
This work utilizes the level 2 IBI product provided by Swarm Alpha and Bravo satellites which can be used to indicate the presence of EPBs \citep{park2013ionospheric}. Since Swarm Alpha and Charlie are closely located, the plasma density observations from both are almost identical. Therefore, this work focuses on Swarm Alpha and Bravo.\par 
According to \citet{park2013ionospheric}, the IBI has three possible values: 0, 1, and -1. Bubble index 0 denotes no small-scale fluctuations are present around that data point, i.e., no EPBs were detected. In contrast, bubble index 1 (bubble) means that EPBs impacted the data point, and bubble index -1 means an unanalyzable IBI event.\par
Similar to the scnLv2 data, the geographical latitude $\in[20\degree,28\degree]$, geographical longitude $\in[47\degree,63\degree]$, AE post-sunset $\in[19,23]$ LT hours, and WS pre-sunset $\in[14,18]$ LT hours were considered.\par

\section{Results \& Discussion}\label{RnD}

\begin{figure}
\centering
\includegraphics[width=0.5\textwidth]{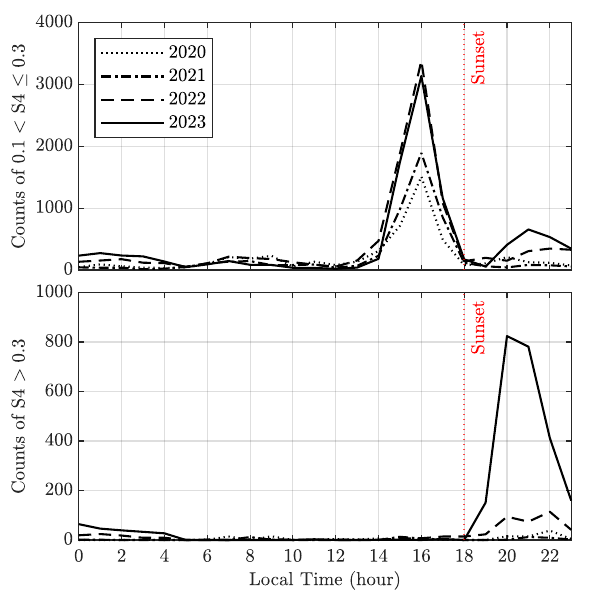}
\caption{Number of observations of $0.1<$ S4 $\leq0.3$ and S4 $>0.3$, respectively, versus local time in hours.}
\label{fig1}
\end{figure}

The diurnal trends of the weak and severe L1 amplitude scintillation are shown in Fig. \ref{fig1} in terms of the number of observations of the 5-minute averaged $0.1<$ S4 $\leq 0.3$ (weak scintillation) and S4 $>0.3$ (strong scintillation) versus LT in hours. The diurnal trend of the count of $0.1<$ S4 $\leq 0.3$ can be seen to increase from $13$ LT, peaking at $16$ LT, then decreasing until $18$ LT. There was also a minor peak of $0.1<$ S4 $\leq 0.3$ at $21$ LT in 2023. On the other hand, the count of S4 $>0.3$ trend can be seen to increase from $18$ LT, peaking at $20$--$22$ LT, then decreasing until $05$ LT. It is noted that the post-midnight occurrences of strong scintillation are considerably less numerous than the post-sunset pre-midnight observations.\par
It is clear from Fig. \ref{fig1} that sunset separates both weak and strong scintillation trends. The peak count of $0.1<$ S4 $\leq 0.3$ was largest in $2022$, followed closely by $2023$. The counts of $0.1<$ S4 $\leq 0.3$ in $2022$ and $2023$ are almost two times those of $2021$ and $2020$. Furthermore, the peak count of S4 $>0.3$ during $2023$ was $7.2\times$ that of $2022$, and $21.7\times$  that of $2021$, which shows considerable enhancement of weak and severe scintillation as a result of higher solar activity. The two main features observed in this figure are the pre-sunset and post-sunset enhancement of weak and strong scintillation, respectively.\par

\begin{figure}
\centering
\includegraphics[width=0.5\textwidth]{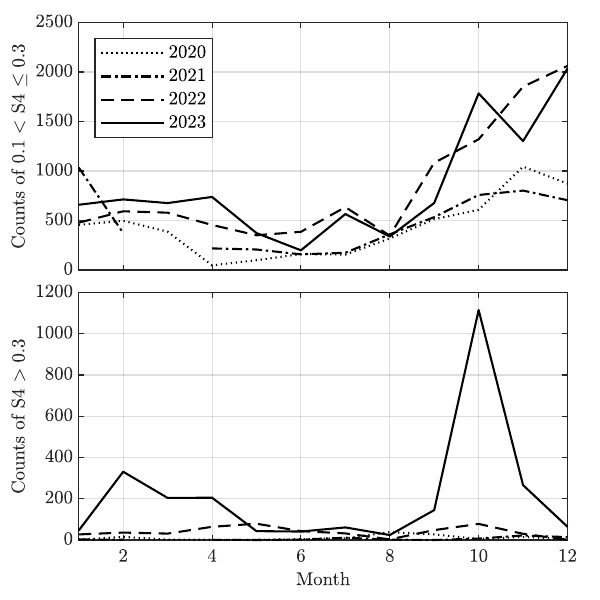}
\caption{Number of observations of $0.1<$ S4 $\leq0.3$ and S4 $>0.3$, respectively, versus months.}
\label{fig2}
\end{figure}

The seasonal trends of the weak ($0.1<$ S4 $\leq 0.3$) and severe (S4 $>0.3$) L1 amplitude scintillation are shown in Fig. \ref{fig2} in terms of the number of observations of 5-minute averaged $0.1<$ S4 $\leq 0.3$ and S4 $>0.3$ versus months. The counts of $0.1<$ S4 $\leq 0.3$ for 2023 and 2022 were highest in December, whereas for 2021 and 2020, the highest count of $0.1<$ S4 $\leq 0.3$ was during November and January, respectively. The lowest counts of $0.1<$ S4 $\leq 0.3$ for 2023, 2022, 2021, and 2020 were during June, August, June, and April, respectively. The counts of S4 $>0.3$ were negligible during 2020, 2021 and 2022, and considerable in 2023. In fact, the peak count of S4 $>0.3$ during 2023 was $14\times$ that of 2022. For 2023, two peak counts of S4 $>0.3$ can be observed, with the highest being in October and the second highest being in February. These two peaks coincide with the autumnal and vernal equinoxes \citep{aarons1971global}. On the other hand, the minimum count of S4 during 2023 was observed in August. The count of S4 $>0.3$ in October was $47\times$ that in August, only two months prior. In summary, weak scintillation was observed during all four years and was more prominent during the winter and less noticeable during the summer. Strong scintillation was mainly observed during 2023, particularly during the autumnal and vernal equinoxes.\par
\begin{figure}
\centering
\begin{minipage}{0.45\textwidth}
  \centering
  \includegraphics[width=0.95\textwidth]{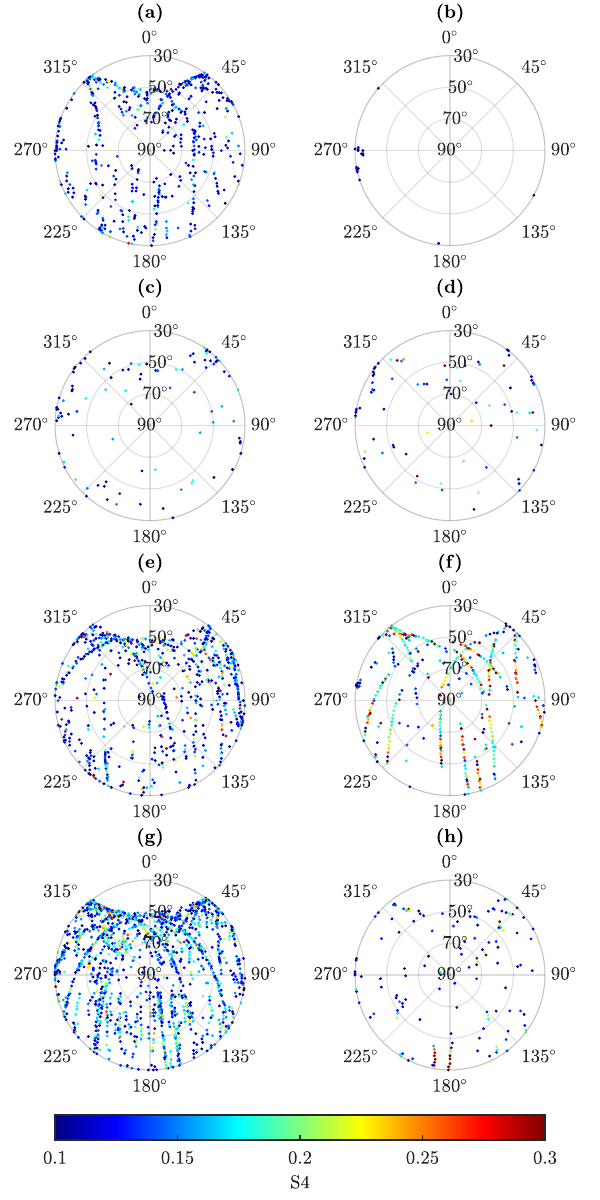}
  \captionof{figure}{These polar plots present the mean S4 observations during 2020 for the following: (a) 14--18 LT Vernal Equinox, (b) 19--23 LT Vernal Equinox, (c) 14--18 LT Summer Solstice, (d) 19--23 LT Summer Solstice, (e) 14--18 LT Autumnal Equinox, (f) 19--23 LT Autumnal Equinox, (g) 14--18 LT Winter Solstice, and (h) 19--23 LT Winter Solstice.}
  \label{fig3}
\end{minipage}
\begin{minipage}{0.1\textwidth}
\end{minipage}
\begin{minipage}{0.45\textwidth}
  \centering
  \includegraphics[width=0.95\textwidth]{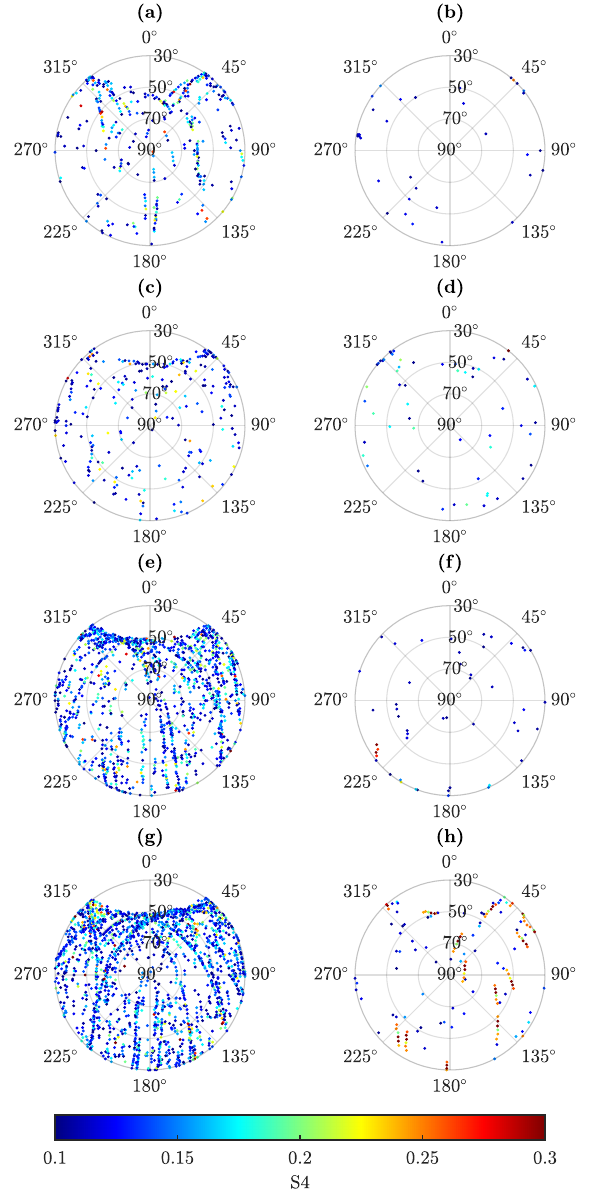}
  \captionof{figure}{These polar plots present the mean S4 observations during 2021 for the following: (a) 14--18 LT Vernal Equinox, (b) 19--23 LT Vernal Equinox, (c) 14--18 LT Summer Solstice, (d) 19--23 LT Summer Solstice, (e) 14--18 LT Autumnal Equinox, (f) 19--23 LT Autumnal Equinox, (g) 14--18 LT Winter Solstice, and (h) 19--23 LT Winter Solstice.}
  \label{fig4}
\end{minipage}
\end{figure}
\begin{figure}
\centering
\begin{minipage}{0.45\textwidth}
  \centering
  \includegraphics[width=0.95\textwidth]{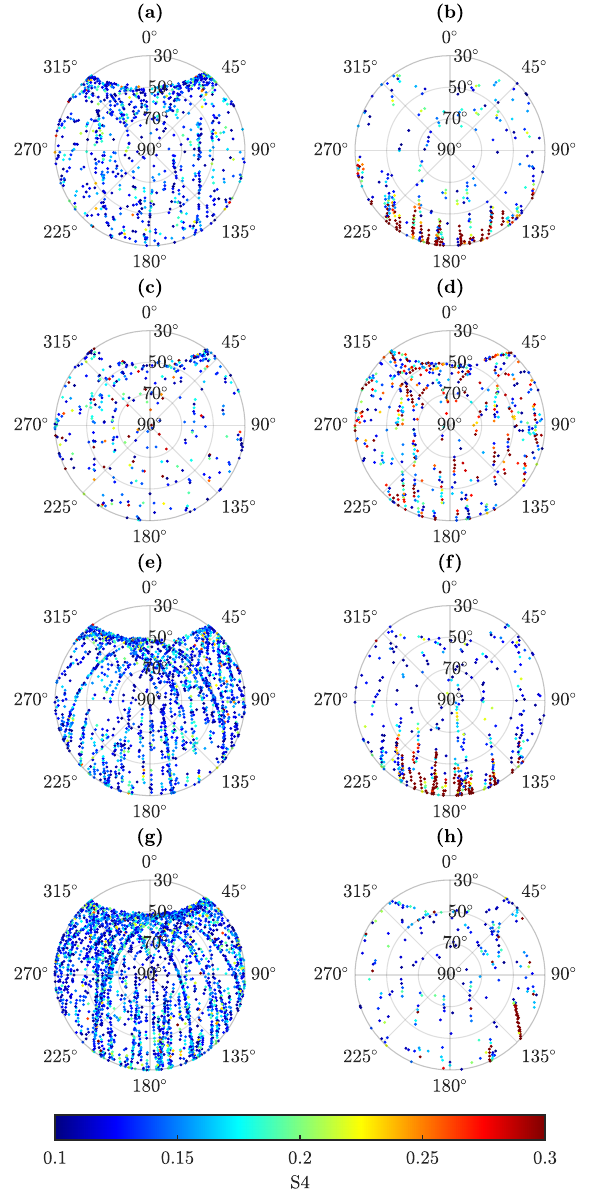}
  \captionof{figure}{These polar plots present the mean S4 observations during 2022 for the following: (a) 14--18 LT Vernal Equinox, (b) 19--23 LT Vernal Equinox, (c) 14--18 LT Summer Solstice, (d) 19--23 LT Summer Solstice, (e) 14--18 LT Autumnal Equinox, (f) 19--23 LT Autumnal Equinox, (g) 14--18 LT Winter Solstice, and (h) 19--23 LT Winter Solstice.}
  \label{fig5}
\end{minipage}
\begin{minipage}{0.1\textwidth}
\end{minipage}
\begin{minipage}{0.45\textwidth}
  \centering
  \includegraphics[width=0.95\textwidth]{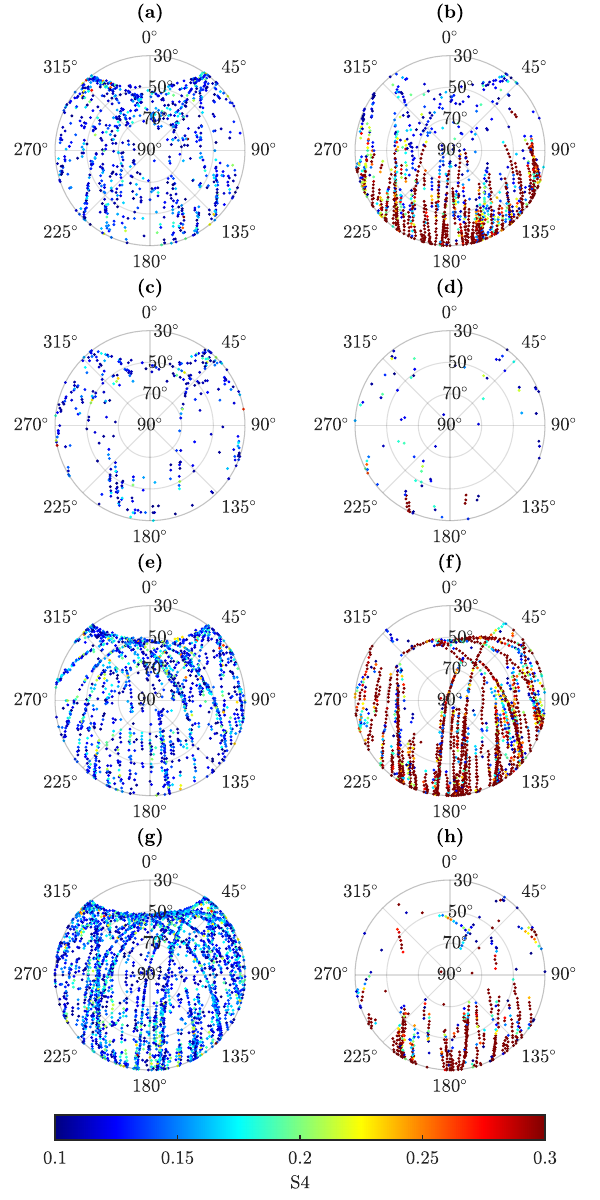}
  \captionof{figure}{These polar plots present the mean S4 observations during 2023 for the following: (a) 14--18 LT Vernal Equinox, (b) 19--23 LT Vernal Equinox, (c) 14--18 LT Summer Solstice, (d) 19--23 LT Summer Solstice, (e) 14--18 LT Autumnal Equinox, (f) 19--23 LT Autumnal Equinox, (g) 14--18 LT Winter Solstice, and (h) 19--23 LT Winter Solstice.}
  \label{fig6}
\end{minipage}
\end{figure}

In addition to the temporal analysis of S4 trends, it is also important to understand the spatial distribution of ionospheric irregularities. A polar plot would be useful in this case as it presents the S4 data in terms of azimuth (angular axis), elevation (radial axis), and intensity (color bar). However, an issue arises regarding overlapping data points caused by consecutive observations from the same azimuth and elevation. To solve this issue, the average intensity of S4 is considered in this work by averaging S4 observations per $1\degree \times 1\degree$ azimuth and elevation. It must be noted that this is in addition to the 5-minute averaging of S4 intensities done previously. Therefore Figs. \ref{fig3}--\ref{fig6} are used in this work to provide a picture of the presence of consistent pre-sunset/post-sunset seasonal patches of small-scale ionospheric irregularities.\par

The rows of Figs. \ref{fig3}--\ref{fig6} represent the seasons, where (a)--(b) represent the VE, (c)--(d) represent the SS, (e)--(f) represent the AE, and (g)--(h) represent the WS. Furthermore, the columns of Figs. \ref{fig3}--\ref{fig6} represent the time period, where the first column represents the pre-sunset period (LT $\in[14,18]$ hours), and the second column represents the post-sunset period (LT $\in[19,23]$ hours). Additionally, Figs. \ref{fig3}--\ref{fig6} present the observations of the years 2020, 2021, 2022, and 2023, respectively.\par

The pre-sunset plots (first column) in Figs. \ref{fig3}--\ref{fig6} show the presence of widespread ionospheric irregularities, particularly during the WS and AE (last two rows). These irregularities induce weak scintillation since the mean value of S4 is observed to be between 0.1--0.2. These pre-sunset irregularities are considerably limited during the VE and SS. These results from the ascending phase of solar cycle 25 mirror those observed from the same region during the solar minimum between solar cycles 24 and 25 \citep{darya2022mapping}. However, these pre-sunset irregularities are more prolific during the 2023 and 2022 WS than the 2021 and 2020 WS, which is caused by the increased solar activity. These observations agree with the seasonal trends seen in Fig. \ref{fig2}.\par

The post-sunset plots (second column) in Figs. \ref{fig3}--\ref{fig6} show a completely different picture to the pre-sunset plots. First, strong-scintillation-causing ionospheric irregularities were observed during the AE and VE of 2022 and the AE, VE, and WS of 2023. These irregularities were not observed in 2021 and 2020, a period of lower solar activity, neither were they observed during a solar minimum period \citep{darya2022mapping}. Second, these irregularities were mainly observed from low to mid elevation towards the south in 2022, and low to high elevation in 2023. The more distributed pattern observed during pre-sunset was not observed post-sunset. Instead, a more localized southward occurrence of strong-scintillation-causing irregularities was observed during the equinoxes. Furthermore, as seen in Fig. \ref{fig2}, these irregularities were more prominent during the AE than the VE, especially in 2023.\par

Two key points can be deduced from Figs. \ref{fig3}--\ref{fig6}. The first is the widespread presence of weak-scintillation-causing pre-sunset ionospheric irregularities with increasing intensity as the solar cycle progresses. These pre-sunset irregularities were more prominent during the WS of 2022 and 2023. The second point is the presence of strong-scintillation-causing post-sunset ionospheric irregularities towards the south, most notably during the AE and VE of 2023. The enhancement of S4 values suggests the presence of small-scale ionospheric irregularities, in the order of a few hundred meters. The weaker pre-sunset scintillation suggests that these irregularities begin to appear in a widely distributed manner a few hours before sunset. These irregularities then dissipate at sunset, followed by the emergence of more dense irregularities southward shortly after sunset. These denser irregularities then dissipate by the early morning hours.\par

In the following paragraphs, various methods and instruments will be used to further investigate two periods, the pre-sunset of WS 2022, with the highest occurrences of weak scintillation, and the post-sunset of AE 2023 with the highest number of observations of strong scintillation.\par

\begin{figure}
\centering
\includegraphics[width=0.5\textwidth]{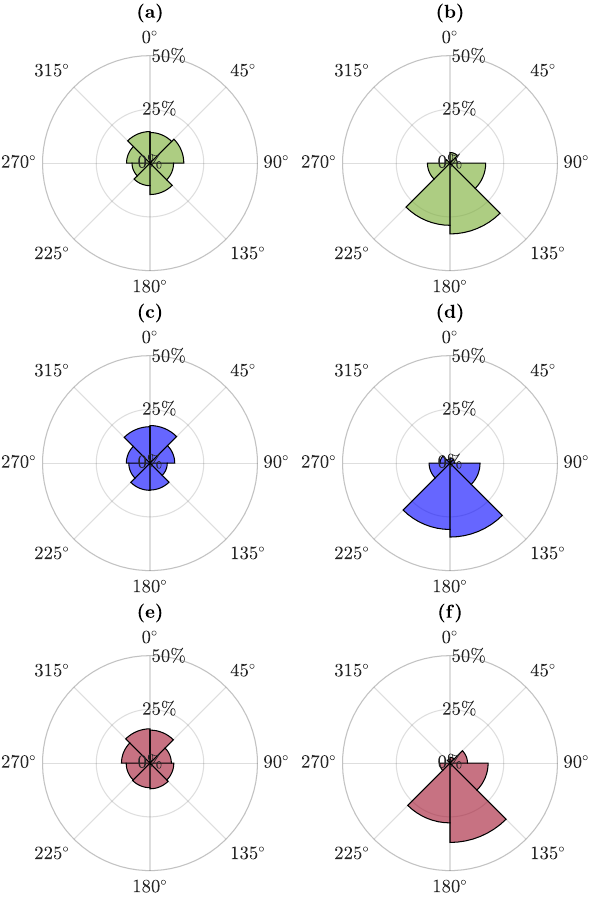}
\caption{This polar histogram represents the S4 observations percentage occurrence per $45\degree$ azimuth for (a) GPS $0.1<$ S4 $\leq0.3$ during 2022, (b) GPS S4 $>0.3$ during 2023, (c) Galileo $0.1<$ S4 $\leq0.3$ during 2022, (d) Galileo S4 $>0.3$ during 2023, (e) BeiDou $0.1<$ S4 $\leq0.3$ during 2022, and (f) BeiDou S4 $>0.3$ during 2023.}
\label{fig7}
\end{figure}

The polar histogram in Fig. \ref{fig7} further analyzes the directional nature of the observed scintillation-causing ionospheric irregularities. The angular axis represents the azimuth, while the radial axis represents the percentage occurrence of $0.1<$ S4 $\leq0.3$ or S4 $>0.3$ per $45\degree$ azimuth. The rows of Fig. \ref{fig7} represent different constellations, where (a)--(b) represent GPS observations, (c)--(d) represent Galileo observations, and (e)--(f) represent BeiDou observations. Additionally, the first column of Fig. \ref{fig7} represents observations of $0.1<$ S4 $\leq0.3$ during 2022 and the second column represents observations of S4 $>0.3$ during 2023.\par

The first column in Fig. \ref{fig7} shows that weak-scintillation-causing ($0.1<$ S4 $\leq0.3$) ionospheric irregularities are distributed across different directions (similar to the first column of Figs. \ref{fig3}--\ref{fig6}), whereas the strong-scintillation-causing (S4 $>0.3$) ionospheric irregularities are concentrated southwards, between 135--225$\degree$, similar to what was observed from the second column of Figs. \ref{fig3}--\ref{fig6}. In fact, the percentage of observations of S4 $>0.3$ observed southwards were $62\%$, $65\%$, and $65\%$ of the total observations for the GPS, Galileo, and BeiDou constellations, respectively. This proves that these observations are constellation-independent since the L1 signals from all three constellations were affected.\par 

\begin{figure}
\centering
\includegraphics[width=0.5\textwidth]{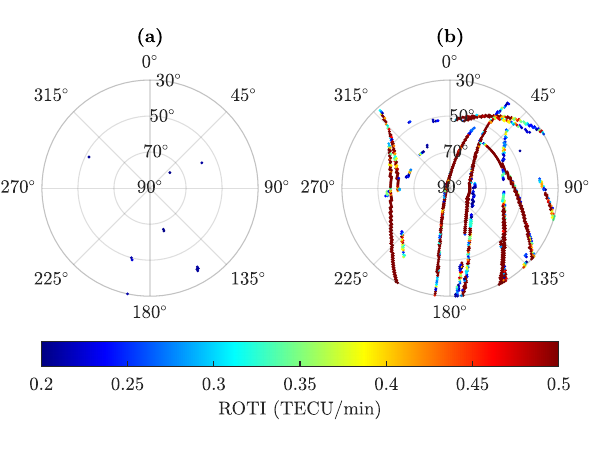}
\caption{These polar plots present the mean ROTI (TECU/min) observations for the following: (a) 14--18 LT 2022 Winter Solstice, (b) 19--23 LT 2023 Autumnal Equinox.}
\label{fig8}
\end{figure}

This work uses a 5-minute averaged ROTI to study the spatial distribution of scintillation-causing ionospheric irregularities with scale sizes in the order of a few kilometers or more (large-scale irregularities) and compare them to the S4 observations of small-scale irregularities in Figs. \ref{fig3}--\ref{fig6}. Fig. \ref{fig8} uses a similar methodology to the one followed in Figs. \ref{fig3}--\ref{fig6}, i.e., using polar plots and averaging ROTI values per $1\degree \times 1\degree$ azimuth and elevation. A difference that must be highlighted, however, is that due to the difference in magnitude between ROTI and S4, the ROTI scale in Fig. \ref{fig8} was set to the range ROTI $=0.2$ (dark blue) to ROTI $\geq 0.5$ (dark red) TECU/minute, since an S4 value of $0.3$ was found to correspond to a ROTI value of $0.5$ in EIA regions \citep{kapil2022reckoning}. Furthermore, unlike Figs. \ref{fig3}--\ref{fig6}, which include multi-constellation observations, the ROTI was derived using GPS observations.\par

Fig. \ref{fig8}(a) shows the 5-minute averaged ROTI observations for the pre-sunset period of WS 2022, while Fig. \ref{fig8}(b) shows the averaged ROTI observations for the post-sunset period of AE 2023. Unlike the pre-sunset WS S4 trend in Fig. \ref{fig5}(g), which shows a wide distribution of weak-scintillation-causing small-scale ionospheric irregularities, the lack of ROTI observations in \ref{fig8}(a) shows that large-scale ionospheric irregularities were not observed during WS 2022. On the other hand, the ROTI trend in post-sunset AE presented in Fig. \ref{fig8}(b) agrees with the S4 trend in Fig. \ref{fig6}(f), with a southward concentration of large-scale irregularities reaching high elevation.\par
By combining the ROTI and S4 observations one may conclude that small-scale ionospheric irregularities are present in both pre- and post-sunset periods, due to the elevated values of S4 in these two periods. Yet, the existence of large-scale irregularities can only be seen during the post-sunset period as evidenced by the elevated ROTI values in that period. Additionally, both large- and small-scale irregularities observed post-sunset were more concentrated towards the south, while the pre-sunset small-scale irregularities were widely distributed.\par

To further study the post-sunset AE period, in addition to ROTI, the TEC anomaly ($\Delta\text{TEC}$) in TECU \citep{nykiel2024large} was produced by band-pass filtering TEC data using a 15--60 minute window. TEC anomalies can be used to detect EPBs or TIDs caused by phenomena such as Acoustic-Gravity Waves.\par

\begin{figure}
\centering
\includegraphics[width=0.5\textwidth]{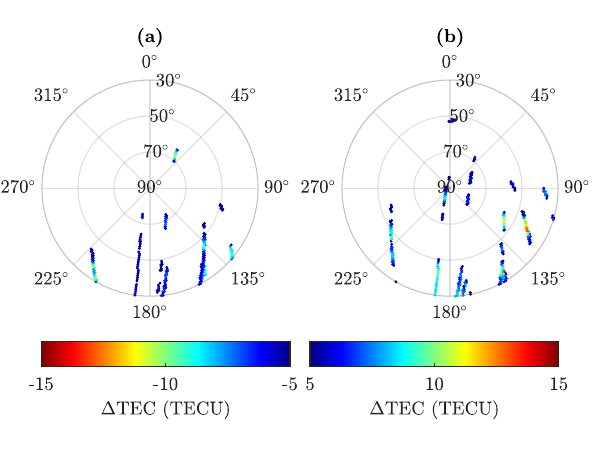}
\caption{These polar plots present $\Delta\text{TEC}$ (TECU) anomalies for 19--23 LT 2023 Autumnal Equinox: (a) negative anomalies, (b) positive anomalies.}
\label{fig9}
\end{figure}

The positive and negative $\Delta\text{TEC}$ anomalies are presented in Fig. \ref{fig9}. All $\Delta\text{TEC}$ anomaly values within $-5<\text{TECU}<5$ were eliminated to focus on the strongest variations of $\Delta\text{TEC}$. All observed $\Delta\text{TEC}$ anomalies in Fig. \ref{fig9} are concentrated southward, with the highest positive anomaly reaching $14$ TECU and the lowest negative anomaly reaching $-11$ TECU. These values suggest the presence of southward TIDs during AE 2023.\par
Using the information provided by the S4, ROTI, and $\Delta\text{TEC}$ analysis, it is clear that strong-scintillation-causing ionospheric irregularities can be detected by all three methods, while weak-scintillation-causing ionospheric irregularities can only be detected using S4 measurements. This is because S4 can detect small-scale ionospheric irregularities, while ROTI and $\Delta\text{TEC}$ can detect large-scale ionospheric irregularities.\par
Spaceborne S4 measurements using RO can complement ground-based measurements and assist in localizing the origin of the ionospheric irregularities in terms of altitude \citep{zou2011ionospheric}. This work utilizes S4 RO data from the F7/C2 mission \citep{chen2021near}. The two periods considered are the pre-sunset WS 2022 and post-sunset AE 2023. The total number of observations of S4 for each of the two periods over the geographical area of interest is listed in Table \ref{tab:table1}. Unlike the ground-based S4 observations of Figs. \ref{fig1}--\ref{fig6}, where the observations of S4 $>0.3$ was greater in post-sunset AE 2023 than in pre-sunset WS 2022, Table \ref{tab:table1} shows that the F7/C2 S4 $>0.3$ observations during pre-sunset WS 2022 was $1.5\times$ that of post-sunset AE 2023. Furthermore, the number of $0.1<$ S4 $\leq 0.3$ observations during post-sunset AE 2023 was $1.3\times$ that of pre-sunset WS 2022, which contradicts the ground-based S4 observations. This may simply be due to the wider coverage of F7/C2 S4 observations, compared to ground-based S4 observations.\par

\begin{table*}
\centering
\caption{Number of scintillation observations from F7/C2.}
\begin{tabular}{|l|l|l|l|}
\hline
Time & Total Num. of Obs. & Counts of $0.1<$ S4 $\leq 0.3$ & Count of S4 $>0.3$ \\
\hline
Pre-sunset WS 2022 & $12\,448$ & $3\,317$ ($26.6\%$) & $695$ ($5.6\%$) \\
\hline
Post-sunset AE 2023 & $12\,809$ & $4\,273$ ($33.4\%$) & $462$ ($3.6\%$) \\
\hline
\end{tabular}
\label{tab:table1}
\end{table*}

\begin{figure}
\centering
\begin{minipage}{0.45\textwidth}
  \centering
  \includegraphics[width=0.95\textwidth]{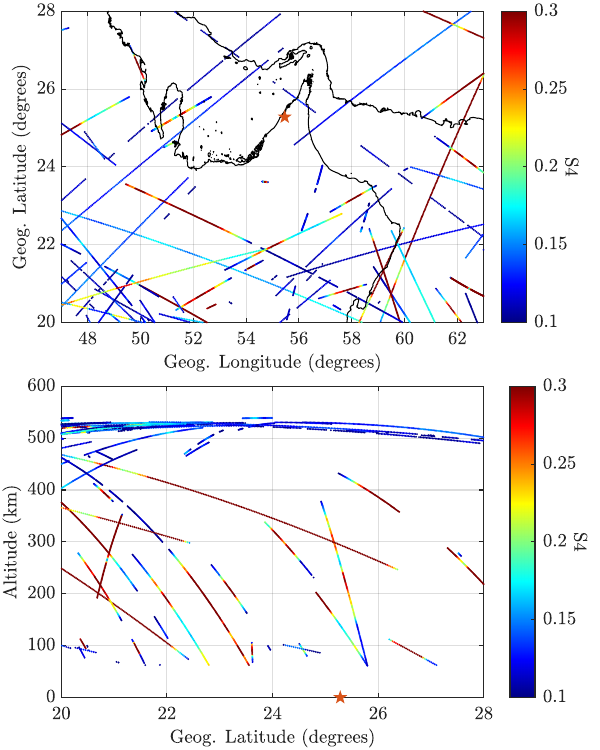}
  \captionof{figure}{The two figures show the spatial distribution of S4 trends taken from the F7/C2 mission for 14--18 LT, Winter Solstice 2022. The red star represents the SHJ1 receiver location. The color bars represent the S4 value ranging from $0.1$ (dark blue) to  $\geq 0.3$ (dark red). The upper figure displays the S4 value in terms of geographical latitude and longitude coordinates. The Lower figure depicts the S4 value in terms of altitude and latitude coordinates.}
  \label{fig10}
\end{minipage}
\begin{minipage}{0.1\textwidth}
\end{minipage}
\begin{minipage}{0.45\textwidth}
  \centering
  \includegraphics[width=0.95\textwidth]{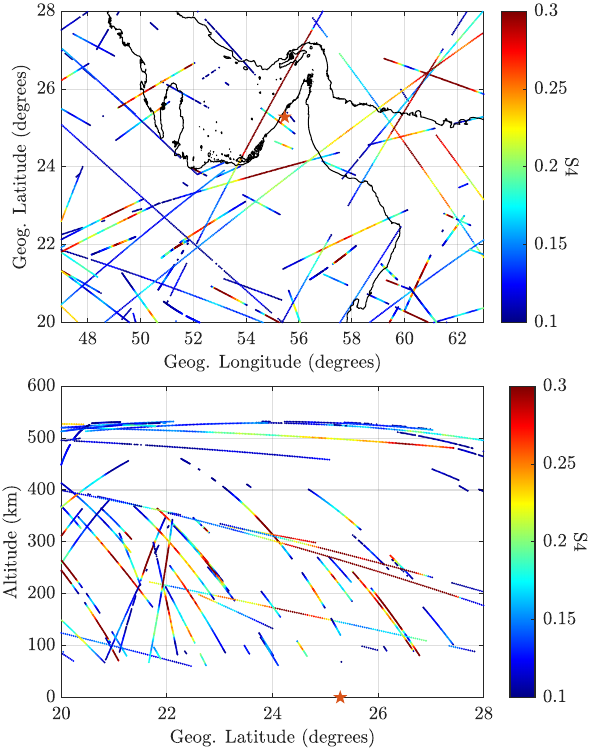}
  \captionof{figure}{The two figures show the spatial distribution of S4 trends observed by the F7/C2 mission for 19--23 LT, Autumnal Equinox 2023. The red star represents the SHJ1 receiver location. The color bars represent the S4 value ranging from $0.1$ (dark blue) to  $\geq 0.3$ (dark red). The upper figure displays the S4 value in terms of geographical latitude and longitude coordinates. The Lower figure depicts the S4 value in terms of altitude and latitude coordinates.}
  \label{fig11}
\end{minipage}
\end{figure}

Figs. \ref{fig10}--\ref{fig11} show the spatial distribution of F7/C2 S4 observations in terms of geographical latitude versus longitude (upper figure), and altitude versus geographical latitude (lower figure). Fig. \ref{fig10} presents the observations for pre-sunset WS of 2022, while Fig. \ref{fig11} presents the observations for post-sunset AE of 2023.\par

In both figures, most of S4 $>0.3$ observations are observed south of the SHJ1 receiver location (represented by a red star in Figs. \ref{fig10}--\ref{fig11}). This agrees with the ground-based S4, ROTI, and $\Delta\text{TEC}$ measurements.\par
\begin{figure}
\centering
\includegraphics[width=0.5\textwidth]{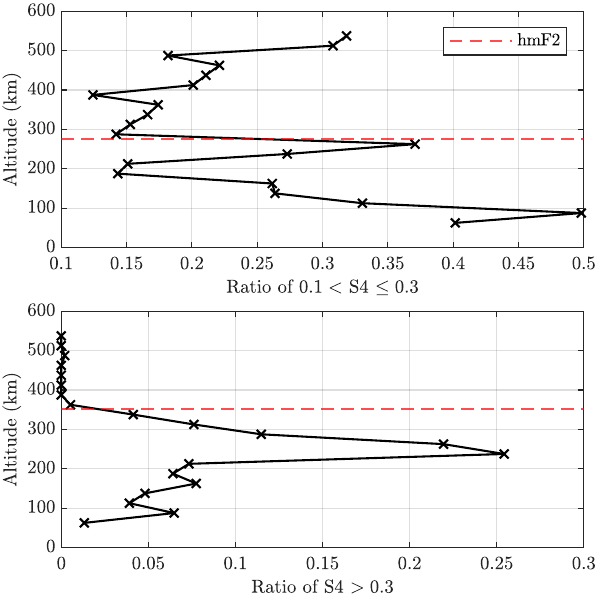}
\caption{This figure shows the ratio of weak and strong scintillation observations per 25 km altitude. The upper figure shows the number of $0.1<$ S4 $\leq0.3$ observations divided by the total number of observations for the 2022 winter solstice. The lower figure shows the number of S4 $>0.3$ observations divided by the total number of observations for the 2023 autumnal equinox. The red dashed line represents the mean hmF2 as observed by DH224.}
\label{fig12}
\end{figure}

What is interesting in Figs. \ref{fig10}--\ref{fig11} is the spatial distribution of the scintillation-inducing irregularities, where observations of S4 $>0.3$ can be seen ranging from as low as 60 km up to 520 km for the pre-sunset WS 2022 and 80 km up to 490 km in post-sunset AE 2023. To better understand how these varying ranges compare with the local ionosphere, Fig. \ref{fig12} shows the ratio of weak and strong scintillation observations per 25 km altitude for pre-sunset WS 2022 (upper figure) and post-sunset AE 2023 (lower figure), in addition to the mean hmF2 value for each period as observed by DH224. These ratios, for each 25 km segment, were obtained by dividing the number of $0.1<$ S4 $\leq0.3$ observations of each segment by the total number of observations for that segment for the upper figure, and dividing the number of S4 $>0.3$ observations by the total number of observations for the lower figure.\par
The upper plot of Fig. \ref{fig12} shows the ratio $0.1<$ S4 $\leq0.3$ exceeding $30\%$ of the total observations at the E-layer (around 100 km), F2-layer, and above the F2-layer. On the other hand, most of the S4 $>0.3$ observations (lower plot of Fig. \ref{fig12}) were under the hmF2, peaking approximately 100 km under the hmF2. From these observations, one may conclude that the pre-sunset WS weak-scintillation-causing irregularities reach above the hmF2, while the post-sunset AE strong-scintillation-causing irregularities remain under the hmF2.\par

Since the majority of strong-scintillation-causing irregularities were observed southward, one of the potential causes is EPBs. The presence of EPBs can be verified using the IBI from Swarm satellites. In Table \ref{tab:table2}, the IBI observations from Swarm Alpha and Bravo are presented. For Swarm Alpha and Bravo, the total number of observations during the pre-sunset WS 2022 and post-sunset AE 2023 periods exceeded $4\,300$ and $6\,600$, respectively. However, the total number of observations where the IBI flag was high, corresponding to the detection of EPBs, was $0$ during the pre-sunset period of WS 2022, and $48$ during the post-sunset period of AE 2023. These $48$ observations were seen during two separate events. The first, with a total of $45$ observations, on 12-Sept-2023 from 20:08:08 to 20:08:52 LT, and a Bubble Probability of $0.25$. The second, with a total of $3$ observations, also on 12-Sept-2023 from 20:09:15 to 20:09:18, with a Bubble Probability of $0$. The Bubble Probability provided by Swarm represents the probability that the magnetic field fluctuation is related to real plasma bubbles. Therefore, due to the low Bubble Probability of these two events it may be concluded that the Swarm satellites were unable to confirm the presence of EPBs over the studied region and duration of interest.\par

\begin{table*}
\centering
\caption{Swarm IBI observations.}
\begin{tabular}{|l|l|l|l|}
\hline
Time & Total Num. of Obs. & Num. of IBI $=1$ Obs. \\
\hline
Pre-sunset WS 2022 & $4\,396$ & $0$ \\
\hline
Post-sunset AE 2023 & $6\,628$ & $48$ \\
\hline
\end{tabular}
\label{tab:table2}
\end{table*}

\section{Conclusions}\label{Conc}
This work used ground and space-based instruments to study the spatial and temporal features of L1 amplitude scintillation-causing ionospheric irregularities over the Eastern Arabian Peninsula region for the ascending phase of solar cycle 25 (years: $2020$--$2023$).\par
The results of the diurnal and seasonal analysis of ground-based S4 showed the presence of pre-sunset and post-sunset enhancements of weak and strong scintillation, respectively. Furthermore, weak scintillation was observed throughout all four years, with it being more prominent during the winter solstice and less noticeable during the summer solstice. Strong scintillation was mainly observed during 2023, particularly during the autumnal and vernal equinoxes.\par
The spatial analysis of the irregularities showed that weaker pre-sunset irregularities began to appear in a widely distributed manner four hours before sunset. These irregularities then dissipated at sunset, followed by the emergence of denser southward irregularities shortly after sunset. These denser irregularities then dissipated by the early morning hours.\par
The combination of ground-based S4 and ROTI observations suggested that small-scale ionospheric irregularities, in the order of a few hundred meters, were present in both pre- and post-sunset periods, due to the elevated values of S4 in these two periods. Yet, the existence of large-scale irregularities, in the order of a few kilometers or more, was only seen during the post-sunset period as evidenced by the elevated ROTI values in that period. Additionally, both large- and small-scale irregularities observed post-sunset were more concentrated towards the south, while the pre-sunset small-scale irregularities were more widely distributed.\par
Spaceborne S4 measurements from the FORMOSAT-7/COSMIC-2 mission showed that strong-scintillation-causing irregularities were more prevalent under the hmF2, while the weak-scintillation-causing irregularities were mostly observed at the E-layer, F2-layer, and above the F2-layer.\par
The presence of southward traveling ionospheric disturbances during the 2023 autumnal equinox was confirmed using ground-based TEC anomaly ($\Delta\text{TEC}$) observations, while the Ionospheric Bubble Index provided by the Swarm mission was unable to confirm the presence of equatorial plasma bubbles during the same period.\par
The findings of this work, in addition to similar studies during the peak of the 25th solar cycle, would shed some light on the behavior of scintillation-causing ionospheric irregularities in this region.\par

\section*{Acknowledgments}
The authors would like to thank the anonymous referees for their helpful comments that improved the quality of the manuscript. The authors would also like to thank Shih-Ping Chen and Manar Anwer Abusirdaneh for their assistance in processing FORMOSAT-7/COSMIC-2 data, and Radhia Fernini for copy-editing the manuscript.\par
The authors acknowledge UCAR for providing FORMOSAT-7/COSMIC-2 data (available at: \url{https://data.cosmic.ucar.edu/}), ESA for providing Swarm data (available at: \url{http://swarm-diss.eo.esa.int/}), and the GIRO group for providing scaled ionosonde data (available at: \url{https://giro.uml.edu/}).\par
Grzegorz Nykiel acknowledges support from the Gdańsk University of Technology through the DEC-31/2021/ IDUB/I.3.3 grant under the ARGENTUM—‘Excellence Initiative— Research University’ program.\par
Essam Ghamry acknowledges ISSI, Bern, Switzerland, and ISSI-BJ, Beijing, China, for the support to the International Team 553 “CSES and Swarm Investigation of the Generation Mechanisms of Low Latitude Pi2 Waves” led by Essam Ghamry and Zeren Zhima and International Team 23-583/57 “Investigation of the Lithosphere Atmosphere Ionosphere Coupling, LAIC, Mechanism before the Natural Hazards” led by Dedalo Marchetti and Essam Ghamry.

\bibliographystyle{jasr-model5-names}
\biboptions{authoryear}
\bibliography{Manuscript.bib}

\end{document}